\documentclass[11pt]{article}
\usepackage{axodraw}
\usepackage{epsfig}
\usepackage{amsfonts}
\usepackage{amsmath}
\usepackage{bbm}
\allowdisplaybreaks[1]

\input pix.sty

\newcommand{\aL}{a^{ }_\rmii{L}}
\newcommand{\aR}{a^{ }_\rmii{R}}
\renewcommand{\eq}{eq.~}
\renewcommand{\eqs}{eqs.~}

\newcommand{\Nc}{N_{\rm c}}

\newcommand{\gammaE}{\gamma_\rmii{E}}

\newcommand{\rmO}{{\mathcal{O}}}
\newcommand{\bmu}{\bar\mu}

\def\lsi{\raise0.3ex\hbox{$<$\kern-0.75em\raise-1.1ex\hbox{$\sim$}}}
\def\gsi{\raise0.3ex\hbox{$>$\kern-0.75em\raise-1.1ex\hbox{$\sim$}}}
\newcommand{\lsim}{\mathop{\lsi}}
\newcommand{\gsim}{\mathop{\gsi}}

\newcommand{\nF}{n_\rmii{F}}

 \renewcommand{\nF}[1]{n_\rmii{F{#1}}}

\newcommand{\rmii}[1]{{\mbox{\tiny\rm{#1}}}}

\newcommand{\re}{\mathop{\mbox{Re}}}
\newcommand{\im}{\mathop{\mbox{Im}}}

\newcommand{\Tint}[1]{{\hbox{$\sum$}\!\!\!\!\!\!\!\int\,}_{\!\!\!\!\raise-0.9ex\hbox{$\scriptstyle{#1}$}}}
\newcommand{\Tinti}[1]{{{\Sigma}\!\!\!\!\raise0.3ex\hbox{$\int$}_\rmii{${#1}$}}}

\newcommand{\bi}{\begin{itemize}}
\newcommand{\ei}{\end{itemize}}

\newcommand{\hide}[1]{ }
\newcommand{\bsl}[1]{\,\slash\!\!\!\!{#1}\,}
\newcommand{\msl}[1]{\,\slash\!\!\!{#1}\,}
\def\TAsc(#1,#2)(#3,#4,#5)%
{\SetWidth{2.0}\CArc(#1,#2)(#3,#4,#5)\SetWidth{1.0}}
\def\Lwidth{3}

\def\TAgl(#1,#2)(#3,#4,#5){\SetWidth{2.0}\PhotonArc(#1,#2)(#3,#4,#5){\Lwidth}%
{6.283 #3 mul 360 div #4 #5 sub #4 #5 sub mul sqrt mul Tdensity mul}%
\SetWidth{1.0}}
\def\TLgl(#1,#2)(#3,#4){\SetWidth{2.0}\Photon(#1,#2)(#3,#4){\Lwidth}
{#1 #3 sub #1 #3 sub mul #2 #4 sub #2 #4 sub mul add sqrt Tdensity mul}%
\SetWidth{1.0}}

\renewcommand{\picc}[1]{\;\parbox[c]{60pt}{\begin{picture}(60,30)(0,0)
\SetWidth{1.0}\SetScale{1.3} #1 \end{picture}}\; }
\def\Lwidth{1.3}

\newcommand{\picu}[1]{\;\parbox[c]{40pt}{\begin{picture}(50,30)(0,0)
\SetWidth{1.0}\SetScale{0.8} #1 \end{picture}}\; }
\def\EleA{\picu{%
 \Asc(30,5)(22.3,27,153)%
 \Aqu(30,25)(22.3,207,333)%
 \COval(10,15)(2,2)(0){Black}{Black}%
 \COval(50,15)(2,2)(0){Black}{Black}%
}}
\def\EleB{\picu{%
 \Asc(30,5)(22.3,27,153)%
 \Aqu(30,25)(22.3,207,333)%
 \COval(10,15)(2,2)(0){Black}{Black}%
 \COval(50,15)(2,2)(0){Black}{Black}%
 \Asc(30,35.0)(8,0,360)%
}}
\def\EleC{\picu{%
 \Asc(30,5)(22.3,27,67)%
 \Asc(30,5)(22.3,113,153)%
 \Aqu(30,25)(22.3,207,333)%
 \COval(10,15)(2,2)(0){Black}{Black}%
 \COval(50,15)(2,2)(0){Black}{Black}%
 \CArc(30,26.3)(9,-90,270)%
 \CArc(30,26.3)(7,-90,270)%
}}
\def\EleD{\picu{%
 \Asc(30,5)(22.3,27,153)%
 \Aqu(30,25)(22.3,207,333)%
 \COval(10,15)(2,2)(0){Black}{Black}%
 \COval(50,15)(2,2)(0){Black}{Black}%
 \Agl(30,35.0)(8,0,360)%
}}
\def\EleE{\picu{%
 \Asc(30,5)(22.3,27,153)%
 \Aqu(30,25)(22.3,207,333)%
 \COval(10,15)(2,2)(0){Black}{Black}%
 \COval(50,15)(2,2)(0){Black}{Black}%
 \Agl(30,26.3)(9,190,350)%
}}
\def\EleF{\picu{%
 \Asc(30,5)(22.3,27,153)%
 \Aqu(30,25)(22.3,207,333)%
 \COval(10,15)(2,2)(0){Black}{Black}%
 \COval(50,15)(2,2)(0){Black}{Black}%
 \Agl(30,3.7)(9,10,170)%
}}
\def\EleG{\picu{%
 \Asc(30,5)(22.3,27,153)%
 \Aqu(30,25)(22.3,207,270)%
 \Aqu(30,25)(22.3,270,333)%
 \COval(10,15)(2,2)(0){Black}{Black}%
 \COval(50,15)(2,2)(0){Black}{Black}%
 \Lgl(30,2.7)(30,27.3)%
}}
\makeatletter \@addtoreset{equation}{section} \makeatother
\renewcommand{\theequation}{\arabic{section}.\arabic{equation}}
\makeatletter
\renewcommand\section{\@startsection {section}{1}{\z@}%
                                   {-5.5ex \@plus -1ex \@minus -.2ex}
                                   {2.3ex \@plus.2ex}%
                                   {\normalfont\large\bfseries}}
\renewcommand\subsection{\@startsection{subsection}{2}{\z@}%
                                     {-3.25ex\@plus -1ex \@minus -.2ex}%
                                     {1.5ex \@plus .2ex}%
                                     {\normalfont\normalsize\bfseries}}
\renewcommand\thesection {\@arabic\c@section}
\renewcommand\thesubsection   {\thesection.\@arabic\c@subsection}
\renewcommand{\@seccntformat}[1]{%
\csname the#1\endcsname.\hspace{1.0em}}
\makeatother

\begin{document}

\flushbottom
\renewcommand{\thefootnote}{\fnsymbol{footnote}}


\begin{flushright}
September 2012 \\
\end{flushright}

\begin{centering}


{\Large{\bf
 Thermal right-handed neutrino self-energy \\[2mm] 
 in the non-relativistic regime\footnote{%
 Presented at the International Workshop {\em  
 Frontiers in Perturbative Quantum Field Theory}, 
 10--12 September 2012, Bielefeld, Germany.
 }
}} 

\vspace{0.5cm}

M.~Laine 

\vspace{0.5cm}

{\em
ITP, AEC, University of Bern, 
Sidlerstrasse 5, CH-3012 Bern, Switzerland\\}



\vspace*{0.8cm}

  
 
\noindent
\small{
\begin{minipage}[c]{14cm}
Recently the issue of radiative corrections to leptogenesis has 
been raised. Considering the ``strong washout'' regime, in which
OPE-techniques permit to streamline the setup, we report the thermal
self-energy matrix of heavy right-handed neutrinos at NLO (resummed
2-loop level) in Standard Model couplings. The renormalized expression
describes flavour transitions and ``inclusive'' decays of chemically
decoupled right-handed neutrinos. Although CP-violation is not
addressed, the result may find use in existing leptogenesis frameworks.
\end{minipage}
}

\end{centering}


 
  



%
\section{Introduction}

Leptogenesis is currently among the most popular scenarios 
for explaining the observed cosmological baryon asymmetry. 
Surprisingly, although the basic mechanism 
is fairly simple~\cite{yanagida}, it appears difficult to develop 
a fully consistent theoretical description of the physics 
involved. The reason is that many different subtle 
topics, such as CP-violation, baryon number violation, 
deviations from thermal equilibrium, 
as well as resummations necessary for a systematic treatment 
of relativistic thermal field theory, 
need all to be put consistently together. 

As far as CP-violation is concerned, it could be both of ``direct'' 
and ``indirect'' type, 
known as ``vertex'' and ``wave function'' corrections,  
respectively~\cite{ls,crv,bp}. The indirect CP-violation 
originating from flavour oscillations is non-trivial 
even at low temperatures where essentially the vacuum formalism
can be used~\cite{fpsw,ap}. Moreover, collective thermal phenomena 
and higher-order corrections could be important~\cite{rr,dl,cp4}. 
At present many approaches remain phenomenological, but 
efforts towards a more systematic treatment are under 
way~\cite{cp0}--\cite{ghk2}.\footnote{%
 The situation is analogous to that for CP-violation in 
 phase transition-based baryogenesis~\cite{pss,clt,hkr}.
} 
Ultimately the goal 
should be to present theoretically consistent results  
in terms of Standard Model couplings, 
in the sense that have recently been obtained  
for CP-conserving rates  
both in the ``ultrarelativistic'' 
($m_\rmi{top} \lsim M \ll \pi T$)~\cite{bb1,bb2} 
and ``non-relativistic'' 
($m_\rmi{top} \lsim \pi T \ll M$)~\cite{salvio,nonrel} 
regimes.
Here we concentrate on the latter regime, called the ``strong washout'' 
case in that memory about initial conditions has been lost.

In efforts towards systematic leptogenesis, 
the right-handed neutrino self-energy plays an important 
role~\cite{abp,ghk1,gh1}. Examples of recent discussions 
can be found in sec.~3 of ref.~\cite{gh2} and  
in sec.~4.1 of ref.~\cite{ghk2}, in both of which the self-energy
was handled at leading order in Standard Model couplings. 
In the temperature regime of interest, 
right-handed neutrinos are out of equilibrium, but the Standard Model
particles are {\em in} equilibrium, at least
as far as CP-conserving reactions are concerned. Therefore
the self-energy of the right-handed neutrinos, which reflects the dynamics
of the other particles, can be computed with established techniques 
of thermal field theory, and there is no reason to restrict
to leading order.  

In this note a result for the thermal right-handed
neutrino self-energy in the non-relativistic regime 
is presented at NLO (partly even NNLO) 
in Standard Model couplings. Compared with our earlier 
work~\cite{nonrel}, where the production rate of right-handed 
neutrinos was computed, the real part of the 
self-energy is added here, 
and the full Lorentz and flavour structures are included. 
Technically, we work at order 
$\displaystyle\rmO(h_\nu^\dagger h_\nu^{ })$
in neutrino Yukawa couplings, thus addressing CP-conserving processes, 
whereas CP-violation originates at the order 
$\displaystyle\rmO(h_\nu^\dagger h_\nu^{ })^2$.
As an outlook, the same observable
could in principle also be computed in 
the ``relativistic'' ($\pi T \sim M$) and ultrarelativistic ($\pi T \gg M$)
regimes; its physics is related to  
``CP even damping and time evolution'',  
as discussed e.g.\ in secs.\ 5.2-3 of ref.~\cite{ms}.

%
\section{Self-energy matrix}

We start by computing the Euclidean correlator
(the Lagrangian and other conventions are summarized
in appendix~A)
\be
 \Sigma_E(K) \equiv 
     \int_0^\beta \! {\rm d}\tau \int_\vec{x} e^{i K\cdot X}
   \Bigl\langle
     (\tilde{\phi}^\dagger \aL \ell)(X) \; 
     (\bar{\ell}\,  \aR \tilde\phi)(0)
   \Bigr\rangle^{ }_T 
 \;, \la{PiE_def}
\ee
where $X = (\tau,\vec{x})$; $K = (k_n,\vec{k})$, 
where $k_n$ are fermionic Matsubara frequencies; 
$\ell$ is a lepton doublet; and $\tilde\phi \equiv i \sigma_2 \phi^*$ 
is a Higgs doublet. The corresponding
retarded correlator is obtained through
the analytic continuation $k_n\to -i [k^0 + i 0^+]$.

Treating $\gamma^5$ in Naive Dimensional Regularization 
and employing Feynman gauge
(other regularizations and gauges 
were discussed in ref.~\cite{nonrel}), we obtain
the contributions (dashed line denotes Higgs, solid lepton, double top, 
wiggly $W^\pm,Z^0$)
\ba
\EleA &=& 
  2\, \Tint{P} \frac{i \aL (\bsl{K} - \bsl{P}) \aR}{P^2(K-P)^2}
 \;, \la{d0} \\[2mm]
\EleB &=& 
  - 12 \lambda_\rmii{B}\,
  \Tint{PQ} \frac{i \aL (\bsl{K} - \bsl{P}) \aR}{Q^2P^4(K-P)^2}\;, \la{d1} \\
\EleC &=& 
  - 2 |h_{t\rmii{B}}|^2 \Nc\,
  \Tint{P\{R\}} \frac{i \aL (\bsl{K} - \bsl{P}) \aR 
  \tr [\aL \bsl{R} \aR (\bsl{P}-\bsl{R}) ]}{P^4(K-P)^2R^2(P-R)^2}\;,
 \la{closed_f}
 \hspace*{10mm} \\
\EleD &=& 
  - (g_{1\rmii{B}}^2 + 3 g_{2\rmii{B}}^2)\,
   \frac{D}{2}\,
  \Tint{PQ} \frac{i \aL (\bsl{K} - \bsl{P}) \aR}{Q^2P^4(K-P)^2}\;, \\
\EleE &=& 
  (g_{1\rmii{B}}^2 + 3 g_{2\rmii{B}}^2)\,
  \frac12\,
  \Tint{PQ} \frac{i \aL (\bsl{K} - \bsl{P}) \aR 
  (P+Q)^2}{P^4Q^2(P-Q)^2(K-P)^2}\;, \\
\EleF &=& 
  (g_{1\rmii{B}}^2 + 3 g_{2\rmii{B}}^2)\,
  \frac12\, 
  \Tint{PQ} \frac{i \aL (\bsl{K} - \bsl{P}) \gamma_\mu
  (\bsl{K} - \bsl{Q}) \gamma_\mu  (\bsl{K} - \bsl{P}) \aR}
  {P^2(P-Q)^2(K-P)^4(K-Q)^2}\;,
  \la{d6} \\
\EleG &=& 
  - (g_{1\rmii{B}}^2 + 3 g_{2\rmii{B}}^2)\,
  \frac12\, 
  \Tint{PQ} \frac{i \aL (\bsl{K} - \bsl{P}) 
  (\bsl{P} + \bsl{Q})  (\bsl{K} - \bsl{Q}) \aR}
 {P^2Q^2(P-Q)^2(K-P)^2(K-Q)^2}\;.
 \la{d7}
\ea
Here $D = 4 - 2\epsilon$; 
$\Tinti{P}$, $\Tinti{Q}$ are bosonic sum-integrals; 
and $\Tinti{\{R\}}$ is a fermionic one.

The sum-integrals in \eqs\nr{d0}--\nr{d7} 
are of a ``tensor'' type~\cite{tensor}; 
the numerator transforms non-trivially in O(4) rotations. It is known 
that in an ultrarelativistic plasma, fermion self-energy does 
not respect Lorentz invariance; the part multiplying $\gamma^0$ 
differs from that multiplying the spatial $\gamma^k$~\cite{haw}.
It turns out, however, that in the non-relativistic regime, 
i.e.\ $(\pi T)^2 \ll K^2$, in which the problem can be discussed
with OPE language~\cite{simon}, 
the first thermal corrections do respect 
Lorentz invariance. Perhaps the simplest way to understand this 
is that thermal corrections of $\rmO(\pi T)^2$ are represented by the 
condensate $\langle \phi^\dagger \phi \rangle^{ }_T$~\cite{nonrel}, 
and there is no room for 
tensor structures in this condensate. 

In practice, the sum-integrals are of two types. 
Those proportional to the external four-momentum $\bsl{K}$ are 
scalars; results can be found in appendix~C of ref.~\cite{nonrel}.
Those proportional to $\bsl{P}$, $\bsl{Q}$ or $\bsl{R}$ are  
tensors; Matsubara sums need to be carried out separately
for the temporal and spatial components. In the OPE regime this can 
be done with the techniques introduced in refs.~\cite{Bulk_OPE,Shear_OPE}.
In each case, only structures proportional to $\bsl{K}$ 
survive at $\rmO(\pi T)^2$.

Concretely, multiplying $\Sigma_E$ by the renormalization factor
related to the neutrino Yukawa couplings, 
$
 \mathcal{Z}_\nu \equiv 
  1 +  \frac{1}{(4\pi)^2\epsilon}
   \bigl[
     |h_t|^2 \Nc - \fr34 (g_1^2 + 3 g_2^2) 
   \bigr]  + \rmO(g^4)
$
where $g$ denotes generic Standard Model couplings
and corrections involving neutrino Yukawa couplings were omitted, 
we obtain 
\ba
 \mathcal{Z}_\nu\, \Sigma_E(K) & = & 
 \aL i \bsl{K} \aR \, \biggl\{ 
 \frac{1}{(4\pi)^2}
 \biggl(
  \frac{1}{\epsilon} + \ln\frac{\bmu^2}{K^2} + 2 
 \biggr) 
 \nn 
 & & \qquad + \, 
 \frac{ |h_t|^2 \Nc }{(4\pi)^4}
 \biggl(
  \frac{1}{2\epsilon^2} - \frac{3}{4\epsilon} 
  - \frac{1}{2} \ln^2\frac{\bmu^2}{K^2} 
  - \frac{7}{2} \ln\frac{\bmu^2}{K^2} 
  - \frac{57}{8}
 \biggr)
 \nn
 & & \qquad + \, 
 \frac{g_1^2 + 3g_2^2}{(4\pi)^4}
 \biggl(
  - \frac{3}{8\epsilon^2} + \frac{17}{16\epsilon} 
  + \frac{3}{8} \ln^2\frac{\bmu^2}{K^2} 
  + \frac{29}{8} \ln\frac{\bmu^2}{K^2} 
  + \frac{275}{32} - 3 \zeta(3)
 \biggr)
 \nn
 & & \qquad + \, 
 \biggl[1 + \frac{6\lambda}{(4\pi)^2}
 \biggl(\ln\frac{\bmu^2}{K^2} + 1 \biggr) \biggr]
 \, \frac{\mathcal{Z}_m \langle \phi^\dagger \phi \rangle^{ }_T}{K^2}
 \; + \, \rmO\Bigl(g^4,\frac{T^4}{K^4}\Bigr) \biggr\} 
 \;. \la{SigmaE}
\ea
Here 
$
 \mathcal{Z}_m \equiv 
 1 + \frac{1}{(4\pi)^2\epsilon}
 \bigl[
   6 \lambda + |h_t|^2\Nc - \fr34 (g_1^2 + 3 g_2^2) 
 \bigr] + \rmO(g^4)
  \la{Zm}
$
is the renormalization factor related to the Higgs mass 
parameter (denoted below by $m_0^2$) and, 
for $\pi T \gsim m_\rmi{top}$~\cite{nonrel},  
\ba
 \mathcal{Z}_m \langle \phi^\dagger \phi \rangle^{ }_T
 &  =  &
  \frac{T^2}{6} - \frac{T^2}{2\pi}
  \sqrt{ \frac{m_\rmii{H}^2}{T^2} - 
 \frac{ g_1^2 m_\rmii{D1}^{} + 3 g_2^2 m_\rmii{D2}^{ }}
  {16\pi T} }
  \nn & & \; + \, 
  \frac{T^2}{48\pi^2}
  \biggl\{
     -6 \lambda \biggl[ \ln\biggl(
   \frac{\bmu e^{\gammaE}}{4\pi T}\biggl)
     - 3 \biggr]
     - |h_t|^2 \Nc \ln\biggl(
      \frac{\bmu e^{\gammaE}}{8\pi T}\biggl)
  \nn & & \; + \, 
  \frac{3(g_1^2 + 3 g_2^2)}{4}
  \biggl[ 
    \ln\biggl(\frac{\bmu e^{\gammaE}}{4\pi T}
    \biggl) - \fr23 - 2 \gammaE - 
    2 \frac{\zeta'(-1)}{\zeta(-1)} + 4 \ln 
   \biggl(\frac{2\pi T}{m_\rmii{H}}\biggl)
  \biggr]
  \biggr\} + \rmO(g^3) 
  \;,  \nn
 \la{pdp} \\
  m_\rmii{H}^2 
 & = & m_0^2 + 
 \biggl(\frac{\lambda}{2} + \frac{|h_t|^2\Nc}{12} + 
 \frac{g_1^2 + 3 g_2^2}{16} \biggr) T^2 + \rmO(g^4)
 \;. 
\ea
Here $m^{ }_\rmii{D1}$, $m^{ }_\rmii{D2}$ are the U$_\rmii{Y}$(1) 
and SU$_\rmii{L}$(2) Debye masses, and chemical potentials have been 
set to zero. All thermal corrections lie in 
$\mathcal{Z}_m \langle \phi^\dagger \phi \rangle^{ }_T$.
{\em If} we count $(\pi T)^2 \sim K^2$ then 
eq.~\nr{SigmaE} is ``NNLO'' because, after analytic
continuation, its real part
contains terms of $\rmO(g^0)$, $\rmO(g^1)$, $\rmO(g^2)$
(cf.\ \eqs\nr{pdp}, \nr{phiR}), 
and its imaginary part terms of 
$\rmO(g^0)$, $\rmO(g^2)$, $\rmO(g^3)$
(cf.\ \eqs\nr{pdp}, \nr{phiI}). However, in the non-relativistic 
regime it is more natural to count $(\pi T)^2 \sim g^2 K^2$, 
and then the imaginary part is complete only up to NLO.

For the next steps, 
a formalism needs to be chosen
by which to represent right-handed neutrinos; various
possibilities are summarized in appendix~A. If the neutrinos 
are represented as chiral Dirac fermions then, 
after the analytic continuation $k_n \to -i [k^0 + i 0 ^+]$, 
\eq\nr{SigmaE} amounts to a ``wave function correction'' 
in their retarded self-energy: 
\be
 \aL \bsl{\mathcal{K}} \aR \to 
 \aL \bsl{\mathcal{K}} \aR
 \Bigl\{  \mathbbm{1} + h_\nu^\dagger h^{ }_\nu 
  \, \bigl[ \phi_\rmii{R}(\mathcal{K}^2) 
 + i \mathop{\mbox{sign}}(k^0)
  \phi_\rmii{I}(\mathcal{K}^2) \bigr] 
 \Bigr\}
 \;, \la{Z_Dirac}
\ee
where $\mathcal{K} = (k^0,\vec{k})$, $h^{ }_\nu$ is 
a $3\times 3$ matrix of Yukawa couplings (cf.\ \eq\nr{L_Dirac}), and 
\ba
 \phi^{ }_\rmii{R}(\mathcal{K}^2) & = & 
 \frac{1}{(4\pi)^2}
 \biggl(
  \frac{1}{\epsilon} + \ln\frac{\bmu^2}{\mathcal{K}^2} + 2 
 \biggr) 
 \nn 
 & + & 
 \frac{ |h_t|^2 \Nc }{(4\pi)^4}
 \biggl(
  \frac{1}{2\epsilon^2} - \frac{3}{4\epsilon} 
  - \frac{1}{2} \ln^2\frac{\bmu^2}{\mathcal{K}^2} 
  - \frac{7}{2} \ln\frac{\bmu^2}{\mathcal{K}^2} 
  - \frac{57}{8} + \frac{\pi^2}{2}
 \biggr)
 \nn
 & + & 
 \frac{g_1^2 + 3g_2^2}{(4\pi)^4}
 \biggl(
  - \frac{3}{8\epsilon^2} + \frac{17}{16\epsilon} 
  + \frac{3}{8} \ln^2\frac{\bmu^2}{\mathcal{K}^2} 
  + \frac{29}{8} \ln\frac{\bmu^2}{\mathcal{K}^2} 
  + \frac{275}{32} - \frac{3\pi^2}{8} - 3 \zeta(3)
 \biggr)
 \nn
 & - & 
 \biggl[1 + \frac{6\lambda}{(4\pi)^2}
 \biggl(\ln\frac{\bmu^2}{\mathcal{K}^2} + 1 \biggr) \biggr]
 \, \frac{\mathcal{Z}_m \langle \phi^\dagger \phi \rangle^{ }_T}{\mathcal{K}^2}
 \, + \, \rmO\Bigl(g^4,\frac{T^4}{\mathcal{K}^4}\Bigr)
 \;, \la{phiR} \\ 
 \frac{\phi^{ }_\rmii{I} (\mathcal{K}^2) }{\pi} & = & 
 \frac{1}{(4\pi)^2}
 - 
 \frac{ |h_t|^2 \Nc }{(4\pi)^4}
 \biggl(
    \ln \frac{\bmu^2}{\mathcal{K}^2} 
  + \frac{7}{2}
 \biggr)
 + 
 \frac{g_1^2 + 3g_2^2}{(4\pi)^4}
 \biggl(
    \frac{3}{4} \ln \frac{\bmu^2}{\mathcal{K}^2} 
  + \frac{29}{8} 
 \biggr)
 \nn
 & - & 
 \frac{6\lambda}{(4\pi)^2}
 \, \frac{\mathcal{Z}_m \langle \phi^\dagger \phi \rangle^{ }_T}{\mathcal{K}^2}
 \, + \, \rmO\Bigl(g^4,\frac{T^4}{\mathcal{K}^4}\Bigr)
 \;. \la{phiI}
\ea
If we rather employ 
Majorana spinors, then \eq\nr{Majorana_kin}
implies that the flavour structure can be ``reflected'' to 
the left-handed components:
\be
 \bsl{\mathcal{K}}  \to 
 \bsl{\mathcal{K}} \; 
 \Bigl\{ \mathbbm{1}  + 
 \bigl[
        h_\nu^\dagger h^{ }_\nu \aR 
   +   (h_\nu^\dagger h^{ }_\nu)^T \aL 
 \bigr]
 \bigl[ 
    \phi_\rmii{R}(\mathcal{K}^2) + 
  i \mathop{\mbox{sign}}(k^0) \phi_\rmii{I}(\mathcal{K}^2) 
 \bigr]
 \Bigr\}
 \;. \la{Z_Majorana}
\ee
Given that $\displaystyle h_\nu^\dagger h^{ }_\nu$ is Hermitean, 
we denote $\displaystyle (h_\nu^\dagger h^{ }_\nu)^T$ by 
$\displaystyle (h_\nu^\dagger h^{ }_\nu)^*$ in the following. 
%
If a time-ordered correlator ($\Sigma^{ }_T$) 
is considered rather than a retarded 
one ($\Sigma^{ }_R$), then the corresponding self-energy reads
\be
 \Sigma^{ }_{T}(\mathcal{K})
 = 
 \re  \Sigma^{ }_{R}(\mathcal{K})
 + 
 i \bigl[ 1 - 2 \nF{}(k^0) \bigr]
 \im  \Sigma^{ }_{R}(\mathcal{K})
 \;,
\ee
where $\nF{}$ denotes the Fermi distribution, 
and $\im$ refers to a cut across the $k^0$-axis. 
In the non-relativistic regime, $|k^0| \gg \pi T$, 
we can simplify
$
 1 - 2 \nF{}(k^0) = \mathop{\mbox{sign}}(k^0)
$,
and therefore in the time-ordered case it is the combination
$
    \phi_\rmii{R}(\mathcal{K}^2) + 
  i \phi_\rmii{I}(\mathcal{K}^2) 
$
that appears. 

%
\section{On-shell renormalization}

As suggested by \eqs\nr{Z_Dirac}, \nr{Z_Majorana}, 
it is conventional to view the divergences appearing in $\phi^{ }_\rmii{R}$, 
\eq\nr{phiR}, as being cancelled by wave function renormalization. 
Wave function normalization being somewhat unphysical, it may be
more elegant to discuss only mass renormalization explicitly. This 
can be achieved by considering renormalization at an on-shell point.  

In the literature, various formalisms are being used for
addressing right-handed neutrino dynamics in the context of leptogenesis. 
Within the ``Kadanoff-Baym''
framework, retarded and advanced self-energies appear directly, 
and the pole positions of the corresponding propagators can 
be solved for (cf.\ e.g.\ sec.~4.1 of ref.~\cite{ghk2}).
Within the ``canonical'' approach, a density matrix between free 
one-particle states is constructed and evolved~\cite{sr,ars,as}. 
In both cases on-shell particles appear and the considerations 
below may apply.

Of course, strictly speaking 
right-handed neutrinos are not on-shell states, 
because they decay. Nevertheless, the width being parametrically
suppressed by $\displaystyle\rmO(h_\nu^{ \dagger} h_\nu^{ })$, 
the concept of an on-shell particle may still be
a useful practical notion, like for the top quark.  
In the following, we refer to the pole mass as the real part of 
the pole position in the $k^0$-plane. 

Let $M^{ }_\rmii{ }$ be the renormalized mass matrix which we choose 
to be non-negative and diagonal, 
and write $M^{ }_\rmii{B} = M^{ }_\rmii{ } + \delta M^{ }_\rmii{ }$. 
The diagonal elements of $M^{ }_\rmii{ }$ are denoted by $m_i$: 
$M = \mathop{\mbox{diag}}(m_1,m_2,m_3)$. Making use of Majorana 
spinors, the effective action for
time-ordered correlators has the form 
\ba
 \mathcal{S}_\rmi{eff} & = & 
 \int_\mathcal{X} \fr12 
 \bar{\tilde{N}}(\mathcal{X})
 \Bigl\{ 
 i \msl{\partial} - M  - 
   \delta M^{ }_\rmii{ } \aR
   -  
   \delta M^{*}_\rmii{ } \aL \nn 
 & &
   + 
   \bigl[
      h_\nu^\dagger h^{ }_\nu \aR + 
      (h_\nu^\dagger h^{ }_\nu)^* \aL
   \bigr]
   i \msl{\partial}
   \bigl[
     \phi^{ }_\rmii{R}(-\partial^2) + i \phi^{ }_\rmii{I}(-\partial^2)  
   \bigr]
 \Bigr\} 
 \tilde{N}(\mathcal{X})
 \;, \quad
 \mathcal{X} \equiv (t,\vec{x})
 \;, \la{Seff1}
\ea
where $\delta M$ denotes the mass counterterm. 
Within the $\displaystyle\rmO(h_\nu^\dagger h_\nu^{ })$ 
part we can make use of tree-level 
equations of motion; denoting 
\be
 \tilde M \equiv \mathop{\mbox{diag}}
 (\tilde m_1,\tilde m_2,\tilde m_3)
 \;, \quad
 \tilde m_i \equiv m_i 
 \bigl[ 
  \phi^{ }_\rmii{R}(m_i^2) + i \phi^{ }_\rmii{I}(m_i^2)
 \bigr]
 \;, 
\ee
and recalling from \eq\nr{L_Weyl} that mass matrices
can always be symmetrized, we get
\ba
 \mathcal{S}_\rmi{eff} & = & 
 \int_\mathcal{X} \fr12 
 \bar{\tilde{N}}(\mathcal{X})
 \Bigl\{ 
 i \msl{\partial} - M  - 
   \delta M^{ }_\rmii{ } \aR
   -  
   \delta M^{*}_\rmii{ } \aL \nn 
 & &
   + 
   \fr12 \bigl[
      h_\nu^\dagger h^{ }_\nu \tilde{M} 
      + \tilde{M}( h_\nu^\dagger h^{ }_\nu)^*
      \bigr] \aR 
   +\fr12 \bigl[  
      (h_\nu^\dagger h^{ }_\nu)^* \tilde{M} 
     + \tilde{M} h_\nu^\dagger h^{ }_\nu 
   \bigr]  \aL
 \Bigr\} 
 \tilde{N}(\mathcal{X})
 \;. \la{Seff2}
\ea
In the on-shell scheme $\delta M$ is chosen to cancel 
the vacuum part of the 
$\displaystyle\rmO(h_\nu^\dagger h_\nu^{ })$ correction; 
hence, writing 
$
\phi_\rmii{R}^\rmii{ } =
\phi_\rmii{R}^\rmii{(0)} +
\phi_\rmii{R}^\rmii{$(T)$}$, 
where $\phi_\rmii{R}^\rmii{(0)}$ corresponds 
to the three first lines of \eq\nr{phiR}, 
\be
 \delta M^{ }_{i_1i_2} \equiv \fr12  
 \Bigl[ 
  ( h_\nu^\dagger h_\nu^{ } )^{ }_{i_1i_2} \, 
   {m^{ }_{i_2} \phi_\rmii{R}^\rmii{(0)}(m_{i_2}^2)
  +
  ( h_\nu^\dagger h_\nu^{ } )^{ }_{i_2i_1} \, 
    m^{ }_{i_1} \phi_\rmii{R}^\rmii{(0)}(m_{i_1}^2)}
 \Bigr]
 \;. \la{ct}
\ee
The finite remainder, which we parametrize through
\be
 \tilde{M}^{ }_r \equiv \mathop{\mbox{diag}}
 (\tilde m^{ }_{r,1},\tilde m^{ }_{r,2},\tilde m^{ }_{r,3})
 \;, \quad
 \tilde m^{ }_{r,i} \equiv m_i \,
 \bigl[ 
  \phi^\rmii{$(T)$}_\rmii{R}(m_i^2) + i 
  \phi^{ }_\rmii{I}(m_i^2)
 \bigr]
 \;, 
\ee
contains thermal ``mass corrections'' 
(from $\phi_\rmii{R}^\rmii{$(T)$}$) as well as a non-Hermitean 
part, representing vacuum and thermal ``decay widths'' 
(from $\phi_\rmii{I}^\rmii{ }$).

To be concrete, consider the canonical formalism 
and define a free Hamiltonian as 
\be
 \hat{\mathcal{H}}_\rmi{eff}^{(0)} (\mathcal{X})
  \equiv
 \fr12\! : \hat{\bar{\tilde{N}}}(\mathcal{X}) 
 \Bigl( 
  - i \gamma^k {\partial}_{k}
  + M 
 \Bigr)
 \hat{\tilde{N}}(\mathcal{X}) :
 \;.
\ee
{}From \eq\nr{Seff2}, 
the first-order correction reads
\be
 \hat{\mathcal{H}}_\rmi{eff}^{(1)} (\mathcal{X})
 - \frac{i}{2} \,
 \hat{ \Gamma }_\rmi{eff}^{(1)} (\mathcal{X})
 \!  \equiv 
 - \fr12 \! : \hat{\bar{\tilde{N}}}(\mathcal{X}) 
 \Bigl( 
   \frac{\mathbbm{1}}{2}
   \bigl\{
      \re (h_\nu^\dagger h^{ }_\nu) ,  \tilde{M}^{ }_r 
   \bigr\}
   + \frac{i\gamma^5}{2}
   \bigl[
      \im (h_\nu^\dagger h^{ }_\nu) , \tilde{M}^{ }_r 
   \bigr]
 \Bigr)
 \hat{\tilde{N}}(\mathcal{X}) : 
 \;. \hspace*{5mm} \la{Heff1}
\ee
Defining a state with a specific flavour ($i_1$), 
momentum ($\vec{k}_1$), and spin ($\tau_1$) formally  as 
\be
 | i_1 \vec{k}_1 \tau_1 \rangle 
 \equiv (2\pi)^{3/2} \hat{a}^\dagger_{i_1 \vec{k}_1 \tau_1 } | 0 \rangle
 \;, 
\ee
and 
inserting the interaction picture field operators from 
\eqs\nr{Nop}, \nr{tNop}, matrix elements can be computed. 
Standard properties of the on-shell spinors 
(cf.\ appendix~A) lead to
\be
 \langle i_1 \vec{k} \tau_1 | 
 \,   \hat{\mathcal{H}}_\rmi{eff}^{(0)}(0)  \, 
 | i_2 \vec{k} \tau_2 \rangle 
 = \delta_{i_1 i_2} \delta_{\tau_1\tau_2} \, \mathcal{K}^{0}_{i_1} 
 \;, \qquad
 \mathcal{K}^{0}_{i_1} = \sqrt{k^2 + m_{i_1}^2}
 \;,
\ee
amounting to a free one-particle energy.
For $i_1 = i_2$, the 
$\displaystyle\rmO(h_\nu^\dagger h_\nu^{ })$
correction yields 
\be
 \langle i_1 \vec{k} \tau_1 | 
 \,   \hat{\mathcal{H}}_\rmi{eff}^{(1)}(0)  \, 
 | i_1 \vec{k} \tau_2 \rangle  = 
 - \frac{m_{i_1}^2}{\mathcal{K}^0_{i_1}} 
 \re(h_\nu^\dagger h_\nu^{ })^{ }_{i_1 i_1} \, 
 \delta^{ }_{\tau_1\tau_2} 
 \phi_\rmii{R}^\rmii{$(T)$}(m_{i_1}^2)
 \;, \la{mass}
\ee
which may be interpreted as a thermal mass correction:
$\delta \mathcal{K}^0_{i_1} \approx
 \displaystyle\re(h_\nu^\dagger h_\nu^{ })^{ }_{i_1 i_1} 
 \mathcal{Z}_m \langle \phi^\dagger \phi \rangle^{ }_T 
 / \mathcal{K}^0_{i_1} 
$. The decay width is 
\ba 
 \langle i_1 \vec{k} \tau_1 | 
 \,   \hat{\Gamma}_\rmi{eff}^{(1)}(0)  \, 
 | i_1 \vec{k} \tau_2 \rangle  & = & 
 \frac{2 m_{i_1}^2}{\mathcal{K}^0_{i_1}} 
 \re(h_\nu^\dagger h_\nu^{ })^{ }_{i_1 i_1} \, 
 \delta^{ }_{\tau_1\tau_2} 
 \phi_\rmii{I}^\rmii{ }(m_{i_1}^2)
 \;, \la{Gamma}
\ea
agreeing with refs.~\cite{salvio,nonrel}. Flavour non-diagonal 
parts are non-zero and can be worked out from
\eqs\nr{bu_u}, \nr{bu_5_u}; in the non-relativistic regime, 
$k \ll m_i$, the spin-part coming from $\gamma^5$ 
is seen to be suppressed, and the flavour structure 
is given by the first term of \eq\nr{Heff1}.

%
\section{Summary}

Addressing the full problem of leptogenesis is 
demanding: although non-perturbative {\em equations} have
been written down,  no general 
systematic {\em solution} is known even to leading order in Standard
Model couplings. However, simpler subproblems can be studied 
in a controlled fashion, and hopefully a similar progress ultimately
permeates the whole topic. 

In this note,  
one simple subproblem has been considered, 
that of 
$\displaystyle\rmO(h_\nu^\dagger h_\nu^{ })$
flavour transitions and 
inclusive decays of heavy 
right-handed neutrinos in the non-relativistic regime
($m_\rmi{top} \lsim \pi T \ll M$). 
In particular, the NLO corrections
to the right-handed neutrino self-energy matrix have been  
determined, cf.\ \eqs\nr{phiR}--\nr{Z_Majorana}. Technically, 
this requires considering ``tensor-type'' sum-integrals~\cite{tensor}. 
Conceivably, generalizing
techniques from ref.~\cite{Bulk_wdep}, a similar computation 
can in the end also be carried out in the relativistic regime.

The main physics conclusion of this study is that in the non-relativistic 
regime thermal effects are (only) power-suppressed, but nevertheless
infrared safe up to the order studied, both as far as mass corrections
and decay widths are
concerned. Whether this statement also applies
to the actual lepton asymmetry generation is, however, 
unclear at present. 
A potential obstacle is that OPE techniques~\cite{simon}, 
which help to understand the structure of the 
current results as well as the nature of higher-order corrections, 
are not trivially applicable. The reason is
that, particularly in the resonant case, 
CP-violating observables contain the heavy mass scale $M$ 
also in internal propagators, 
not only in external states.

%
\section*{Acknowledgements}

I am grateful to Y.~Schr\"oder for collaboration on ref.~\cite{nonrel}
and for helpful discussions. 

%
\appendix
\renewcommand{\thesection}{Appendix~\Alph{section}}
\renewcommand{\thesubsection}{\Alph{section}.\arabic{subsection}}
\renewcommand{\theequation}{\Alph{section}.\arabic{equation}}

%
\section{Conventions for Majorana spinors}
\la{app:A}

For completeness, we specify in this appendix our
conventions for right-handed
neutrinos. As usual, they can be represented by two-component
Weyl spinors ($\xi$), four-component Majorana spinors 
($\tilde{N}$), or right-handed chiral projections of Dirac spinors
($\nu_\rmii{R} \equiv a_\rmii{R}\nu$). 

Choosing the Weyl representation for the Dirac matrices, 
\be
 \gamma^0 \equiv 
 \left( \begin{array}{rr}
   {0} & \mathbbm{1} \\ 
   \mathbbm{1} & {0} 
 \end{array} \right)
 \;, \quad 
 \gamma^k \equiv 
 \left( \begin{array}{rr}
   {0}\; & \sigma_k \\ 
   -\sigma_k & {0}\; 
 \end{array} \right)
 \;, \quad 
 \gamma^5 = 
 \left( \begin{array}{rr}
   -\mathbbm{1} & {0}   \\ 
   {0} & \mathbbm{1} 
 \end{array} \right)
 \;,
\ee
where $\sigma^{ }_k$ are the Pauli matrices, 
the chiral projectors $a_\rmii{L,R} \equiv (\mathbbm{1} \mp \gamma^5)/2$
become diagonal, and the four-component Dirac spinor can be decomposed as 
\be
 \nu = (a_\rmii{L} + a_\rmii{R}) \nu = \nu_\rmii{L} + \nu_\rmii{R} \;, \quad
 \nu_\rmii{R} \equiv 
 \left( \begin{array}{l} 0 \\ \xi \end{array} \right)
 \;.
\ee
The charge conjugation matrix can be defined to be 
\be
 C \;\equiv\; i \gamma^2 \gamma^0 = 
 \left( \begin{array}{rr}
   i \sigma_2 & 0 \\ 
   0 & - i \sigma_2 
 \end{array} \right)
 \;, 
\ee
and if we set
\be
 \tilde N \equiv 
  \left( \begin{array}{c}
   i \sigma_2 \xi^* \\ 
   \xi 
 \end{array} \right) 
 \;, \quad
  \bar{\tilde{N}} = 
  \left( 
  \xi^\dagger \; - \xi^T i \sigma^{ }_2
 \right) 
 \;, \la{def_N}
\ee
then it is easy to see that
$
 \tilde{N}^c \; \equiv \; C \bar{\tilde{N}}^T = \tilde{N}
$, 
i.e.\ that $\tilde{N}$ is a Majorana spinor. 

In terms of Dirac spinors, the bare Lagrangian of the 
right-handed neutrinos reads
\be
 \mathcal{L} \equiv 
 \bar{\nu}^{ }_\rmii{R} i \gamma^\mu\partial_\mu \nu^{ }_\rmii{R}
 - 
 \Bigl(
   \bar{\ell}\,  \aR \tilde\phi\, h^{ }_{\nu\rmii{B}}\, \nu^{ }_\rmii{R} 
 + \fr12 \bar{\nu}_\rmii{R}^c M^{ }_\rmii{B} \nu^{ }_\rmii{R} 
 + \mbox{H.c.}
 \Bigr)
 \;. \la{L_Dirac}
\ee
Issues related to $\gamma^5$ were briefly 
recalled in ref.~\cite{nonrel}.
Both $h^{ }_{\nu\rmii{B}}$ and $M^{ }_\rmii{B}$ 
are matrices in flavour space; 
the Grassmann nature of the fields and the antisymmetry 
of $C$ imply that the matrix $M^{ }_\rmii{B}$ is symmetric, 
$M_\rmii{B}^T = M_\rmii{B}^{ }$. Making use of Weyl spinors, 
the Lagrangian can be rewritten as 
\be
 \mathcal{L} = 
 \xi^\dagger i \bar{\sigma}^\mu \partial_\mu \xi 
 - 
 \Bigl(
   {\ell}_\rmii{L}^\dagger \,  \tilde\phi\, h^{ }_{\nu\rmii{B}}\, \xi 
 - \fr12 \xi^T \, i \sigma^{ }_2 \, M^{ }_\rmii{B}\, \xi  
 + \mbox{H.c.}
 \Bigr)
 \;, \la{L_Weyl} 
\ee
where $\bar{\sigma}^\mu \equiv (\mathbbm{1},\sigma_k)$, 
and $\ell_\rmii{L}$ is to be interpreted as a 2-component spinor. 
Finally, for the Majorana representation, \eq\nr{def_N} implies that 
\ba
 \bar{\tilde{N}} i \gamma^\mu \partial_\mu \aR \tilde{N} 
 & = & 
 \xi^\dagger i \bar{\sigma}^\mu \partial_\mu \xi
 \;, \quad \quad
 \bar{\tilde{N}} i \gamma^\mu \partial_\mu \aL \tilde{N} 
 \;\; = \;\; 
 \xi^T i \bar{\sigma}^{\mu T} \partial_\mu \xi^*  
 \la{Majorana_kin}
 \;, \\ 
  \bar{\tilde{N}} M^{ }_\rmii{B} \aR \tilde{N}
 & = & 
 -\xi^T \, i \sigma^{ }_2 \, M^{ }_\rmii{B} \, \xi
 \;, \quad
  \bar{\tilde{N}} M^{*}_\rmii{B} \aL \tilde{N}
 \;\;  =  \;\; 
   \xi^\dagger \, i \sigma^{ }_2 \, M^{ * }_\rmii{B} \, \xi^* 
 \;.
\ea
Omitting a total derivative from the kinetic term
(or symmetrizing that in \eq\nr{L_Weyl}), 
\be
 \mathcal{L} = 
 \fr12 \bar{\tilde{N}} ( i \gamma^\mu\partial_\mu 
 - M^{ }_\rmii{B} \aR - M^{* }_\rmii{B} \aL ) \tilde{N}
 - 
 \Bigl(
   \bar{\ell}\,  \aR \tilde\phi\, h^{ }_{\nu\rmii{B}}\, \tilde{N} 
 + \mbox{H.c.}
 \Bigr)
 \;. \la{L_Majorana}
\ee

A pleasant feature of the Majorana formulation is that the free 
on-shell condition has the form of the usual Dirac equation, 
$
 ( i \gamma^\mu\partial_\mu  
 - M^{ }_\rmii{B} \aR - M^{ * }_\rmii{B} \aL ) \tilde{N} = 0
$.
If the mass matrix is diagonal, with eigenvalues $m_i \in \mathbbm{R}^+$, then 
field operators of the interaction picture can be expanded as
\ba
 \hat{\tilde{N}}_i(\mathcal{X})
 & = &  
 \int\!\frac{{\rm d}^3\vec{k}}{\sqrt{(2\pi)^3 2 \mathcal{K}^0_i}}
 \sum_\tau 
 \Bigl(
        u^{ }_{i\vec{k}\tau} 
  \hat{a}^{ }_{i\vec{k}\tau} 
        e^{- i \mathcal{K}_i\cdot\mathcal{X}}
 +  
        v^{ }_{i\vec{k}\tau} 
  \hat{a}^{\dagger }_{i\vec{k}\tau} 
        e^{i \mathcal{K}_i\cdot\mathcal{X}} 
 \Bigr)
 \;, \la{Nop} \\ 
 \hat{\bar{\tilde{N}}}_i(\mathcal{X})
 & = &  
 \int\!\frac{{\rm d}^3\vec{k}}{\sqrt{(2\pi)^3 2 \mathcal{K}^0_i}}
 \sum_\tau 
 \Bigl(
        \bar{u}^{ }_{i\vec{k}\tau} 
  \hat{a}^{ \dagger }_{i\vec{k}\tau} 
        e^{i \mathcal{K}_i\cdot\mathcal{X}}
 +  
        \bar{v}^{ }_{i\vec{k}\tau} 
  \hat{a}^{ }_{i\vec{k}\tau} 
        e^{ - i \mathcal{K}_i\cdot\mathcal{X}} 
 \Bigr)
 \;, \la{tNop}
\ea
where $v^{ }_{i\vec{k}\tau} \equiv C \bar{u}^{T}_{i\vec{k}\tau}$ and 
$\bar{v}^{ }_{i\vec{k}\tau} = {u}^{T}_{i\vec{k}\tau} C$; $\tau$ enumerates
the spin states; and $\mathcal{K}^0_i \equiv \sqrt{k^2 + m_i^2}$. 
The creation and annihilation operators are assumed to satisfy
\be
 \{ 
  \hat{a}^{ }_{i_1\vec{k}_1\tau_1} , 
  \hat{a}^{\dagger }_{i_2\vec{k}_2\tau_2}
 \} 
 = \delta^{ }_{i_1i_2} \delta^{ }_{\tau_1\tau_2} 
 \delta^{(3)}(\vec{k}_1 - \vec{k}_2) 
 \;, \la{quant} 
\ee
and the on-shell spinors obey 
$
  (\bsl{\mathcal{K}}_{\!i} - m_i) u^{ }_{i\vec{k}\tau} = 
  (\bsl{\mathcal{K}}_{\!i} + m_i) v^{ }_{i\vec{k}\tau} = 0 
$.
Writing
\be
     u^{ }_{i\vec{k}\tau} 
     \equiv  \frac{\bsl{\mathcal{K}}_{\!i} + m_i}
     {\sqrt{\mathcal{K}^0_i + m_i}} 
     \, \eta_\tau 
     \;, \quad
    \eta^{ }_{\tau} \equiv \frac{1}{\sqrt{2}}
  \left( 
   \begin{array}{c}
     |\tau\rangle \\ |\tau\rangle  
   \end{array}
  \right)
  \;, \la{rep}     
\ee
where 
$
 |\tau\rangle
$
are eigenstates of $\hat{\vec{n}}\cdot\sigma$ with $\hat{\vec{n}}$
an arbitrary unit vector, 
on-shell spinors obey the same relations as in the Dirac case: 
\ba
 & & 
 - \bar{v}^{ }_{i\vec{k}\tau_2}
 v^{ }_{i\vec{k}\tau_1}
 = 
 \bar{u}^{ }_{i\vec{k}\tau_1}
 u^{ }_{i\vec{k}\tau_2}
 \; =  \; 2 m_i \, \delta^{ }_{\tau_1\tau_2} 
 \;, \quad
 \bar{u}^{ }_{i\vec{k}\tau_1}
 v^{ }_{i\vec{k}\tau_2} = 
 \bar{v}^{ }_{i\vec{k}\tau_1}
 u^{ }_{i\vec{k}\tau_2} = 
 0
 \;, \\  
 & & 
 \bar{u}^{ }_{i\vec{k}\tau_1}
 \gamma^0 
 u^{ }_{i\vec{k}\tau_2} = 
 \bar{v}^{ }_{i\vec{k}\tau_1}
 \gamma^0 
 v^{ }_{i\vec{k}\tau_2} 
 \; = \; 
 2 \mathcal{K}^0_i \delta^{ }_{\tau_1\tau_2} 
 \;, \quad
 \bar{u}^{ }_{i\vec{k}\tau_1}
 \gamma^0 
 v^{ }_{i-\vec{k}\tau_2} = 
 \bar{v}^{ }_{i\vec{k}\tau_1}
 \gamma^0 
 u^{ }_{i-\vec{k}\tau_2} = 
 0
 \;. \hspace*{1cm} 
\ea
Making use of 
$
 \sum_\tau \eta^{ }_\tau \bar{\eta}_\tau = \fr12 (\mathbbm{1} + \gamma^0 )
$,  
the usual completeness relations are readily verified:
$
 \sum_\tau  
  u^{ }_{i\vec{k}\tau}
 \bar{u}^{ }_{i\vec{k}\tau} = \bsl{\mathcal{K}}_{\!i} + \mathbbm{1} m_i
 \;, \quad
 \sum_\tau  
  v^{ }_{i\vec{k}\tau}
 \bar{v}^{ }_{i\vec{k}\tau} = \bsl{\mathcal{K}}_{\!i} - \mathbbm{1} m_i
$.
For different flavours, the representation 
of \eq\nr{rep}
can be used for showing that 
\ba
 \bar{u}^{ }_{i_1\vec{k}\tau_1}
 \; {u}^{ }_{i_2\vec{k}\tau_2}
 & = &  
 \delta^{ }_{\tau_1\tau_2} \, 
 \Biggl\{
 \frac{(\mathcal{K}^0_{i_1} + m_{i_1})(\mathcal{K}^0_{i_2} + m_{i_2})
 - k^2}{\sqrt{\mathcal{K}^0_{i_1} + m_{i_1}}
 \sqrt{\mathcal{K}^0_{i_2} + m_{i_2}} }
 \Biggr\}
 \;, \la{bu_u} \\
  \bar{u}^{ }_{i_1\vec{k}\tau_1}
 \,\gamma^5\, {u}^{ }_{i_2\vec{k}\tau_2}
 & = &  
 \langle \tau_1 | \vec{\sigma}\cdot\vec{k} | \tau_2\rangle \, 
 \Biggl\{ 
 \sqrt{ 
 \frac{\mathcal{K}^0_{i_2} + m_{i_2}}{\mathcal{K}^0_{i_1} + m_{i_1}}
 } - 
 \sqrt{ 
 \frac{\mathcal{K}^0_{i_1} + m_{i_1}}{\mathcal{K}^0_{i_2} + m_{i_2}}
 }
 \Biggr\}
 \;. \la{bu_5_u}
\ea



\begin{thebibliography}{99}

\bibitem{yanagida}
  M.~Fukugita and T.~Yanagida,
  {\em Baryogenesis Without Grand Unification,}
  Phys.\ Lett.\  B {174} (1986) 45.

\bibitem{ls}
  J.~Liu and G.~Segr\`e,
  {\em Re-examination of generation of baryon and lepton number
  asymmetries by heavy particle decay,}
  Phys.\ Rev.\ D {48} (1993) 4609
  [hep-ph/9304241].

\bibitem{crv}
  L.~Covi, E.~Roulet and F.~Vissani,
  {\em CP violating decays in leptogenesis scenarios,}
  Phys.\ Lett.\ B {384} (1996) 169
  [hep-ph/9605319].

\bibitem{bp}
  W.~Buchm\"uller and M.~Pl\"umacher,
  {\em CP asymmetry in Majorana neutrino decays,}
  Phys.\ Lett.\ B {431} (1998) 354
  [hep-ph/9710460].

\bibitem{fpsw}
  M.~Flanz, E.A.~Paschos, U.~Sarkar and J.~Weiss,
  {\em Baryogenesis through mixing of heavy Majorana neutrinos,}
  Phys.\ Lett.\ B {389} (1996) 693
  [hep-ph/9607310].

\bibitem{ap}
  A.~Pilaftsis,
  {\em CP violation and baryogenesis due to heavy Majorana neutrinos,}
  Phys.\ Rev.\ D {56} (1997) 5431
  [hep-ph/9707235].

\bibitem{rr}
  G.F.~Giudice, A.~Notari, M.~Raidal, A.~Riotto and A.~Strumia,
  {\em Towards a complete theory of thermal leptogenesis in the SM and MSSM,}
  Nucl.\ Phys.\ B {685} (2004) 89
  [hep-ph/0310123].

\bibitem{dl}
  A.~Abada, S.~Davidson, F.-X.~Josse-Michaux, M.~Losada and A.~Riotto,
  {\em Flavor issues in leptogenesis,}
  JCAP {04} (2006) 004
  [hep-ph/0601083].

\bibitem{cp4}
  C.S.~Fong, M.C.~Gonzalez-Garcia and J.~Racker,
  {\em CP Violation from Scatterings with Gauge Bosons in Leptogenesis,}
  Phys.\ Lett.\  {B 697} (2011)  463
  [1010.2209].

\bibitem{cp0}
  J.-S.~Gagnon and M.~Shaposhnikov,
  {\em Baryon Asymmetry of the Universe without
  Boltzmann or Kadanoff-Baym equations,}
  Phys.\ Rev.\  {D 83} (2011)  065021
  [1012.1126].

\bibitem{cp1}
  A.~Anisimov, W.~Buchm\"uller, M.~Drewes and S.~Mendizabal,
  {\em Quantum Leptogenesis I,}
  Annals Phys.\  {326 } (2011)  1998
  [1012.5821].

\bibitem{cpn}
  C.~Kiessig and M.~Pl\"umacher,
  {\em Hard-Thermal-Loop Corrections in Leptogenesis I: CP-Asymmetries,}  
  JCAP {07} (2012) 014
  [1111.1231].

\bibitem{gh2}
  B.~Garbrecht and M.~Herranen,
  {\em Effective Theory of Resonant Leptogenesis  
  in the Closed-Time-Path Approach,}
  Nucl.\ Phys.\ B {861} (2012) 17
  [1112.5954].

\bibitem{ghk2}
  M.~Garny, A.~Kartavtsev and A.~Hohenegger,
  {\em Leptogenesis from first principles in the resonant regime,}
  1112.6428.

\bibitem{pss}
  T.~Konstandin, T.~Prokopec, M.G.~Schmidt and M.~Seco,
  {\em MSSM electroweak baryogenesis and flavor mixing in transport equations,}
  Nucl.\ Phys.\ B {738} (2006) 1
  [hep-ph/0505103].


\bibitem{clt}
  V.~Cirigliano, C.~Lee and S.~Tulin,
  {\em Resonant Flavor Oscillations in Electroweak Baryogenesis,}
  Phys.\ Rev.\ D {84} (2011) 056006
  [1106.0747].

\bibitem{hkr}
  C.~Fidler, M.~Herranen, K.~Kainulainen and P.M.~Rahkila,
  {\em Flavoured quantum Boltzmann equations from cQPA,}
  JHEP {02} (2012) 065
  [1108.2309].

\bibitem{bb1}
  A.~Anisimov, D.~Besak and D.~B\"odeker,
  {\it Thermal production of relativistic Majorana neutrinos: 
  Strong enhancement by multiple soft scattering,}
  JCAP {03} (2011) 042
  [1012.3784].

\bibitem{bb2}
  D.~Besak and D.~B\"odeker,
  {\em Thermal production of ultrarelativistic right-handed 
  neutrinos: Complete leading-order results,}
  JCAP {03} (2012) 029
  [1202.1288].

\bibitem{salvio}
  A.~Salvio, P.~Lodone and A.~Strumia,
  {\it Towards leptogenesis at NLO: 
  the right-handed neutrino interaction rate,}
  JHEP {08} (2011) 116
  [1106.2814].

\bibitem{nonrel}
  M.~Laine and Y.~Schr\"oder,
  {\em Thermal right-handed neutrino production rate
  in the non-relativistic regime,}
  JHEP {02} (2012) 068
  [1112.1205].

\bibitem{abp}
  A.~Anisimov, A.~Broncano and M.~Pl\"umacher,
  {\em The CP-asymmetry in resonant leptogenesis,}
  Nucl.\ Phys.\ B {737} (2006) 176
  [hep-ph/0511248].

\bibitem{ghk1}
  M.~Garny, A.~Hohenegger, A.~Kartavtsev and M.~Lindner,
  {\em Systematic approach to leptogenesis in nonequilibrium QFT: 
  Self-energy contribution to the CP-violating parameter,}
  Phys.\ Rev.\ D {81} (2010) 085027
  [0911.4122].

\bibitem{gh1}
  M.~Beneke, B.~Garbrecht, C.~Fidler, M.~Herranen and P.~Schwaller,
  {\em Flavoured Leptogenesis in the CTP Formalism,}
  Nucl.\ Phys.\ B {843} (2011) 177
  [1007.4783].

\bibitem{ms}
  M.~Shaposhnikov,
  {\em The $\nu$MSM, leptonic asymmetries, and properties of singlet fermions,}
  JHEP {08} (2008) 008
  [0804.4542].

\bibitem{tensor}
  I.~Ghisoiu and Y.~Schr\"oder,
  {\em A new method for taming tensor sum-integrals,}
  1208.0284.

\bibitem{haw}
  H.A.~Weldon,
  {\it Effective fermion masses of $\rmO(gT)$ 
  in high-temperature gauge theories
  with exact chiral invariance,}
  Phys.\ Rev.\ D {26} (1982) 2789.

\bibitem{simon}
  S.~Caron-Huot,
  {\it Asymptotics of thermal spectral functions,}
  Phys.\ Rev.\  D {79} (2009) 125009
  [0903.3958].

\bibitem{Bulk_OPE}
  M.~Laine, M.~Veps\"al\"ainen and A.~Vuorinen,
  {\it Ultraviolet asymptotics of scalar and pseudoscalar
  correlators in hot Yang-Mills theory,}
  JHEP {10} (2010) 010
  [1008.3263].

\bibitem{Shear_OPE}
  Y.~Schr\"oder, M.~Veps\"al\"ainen, A.~Vuorinen and Y.~Zhu,
  {\em The Ultraviolet limit and sum rule for the shear correlator 
  in hot Yang-Mills theory,}
  JHEP {12} (2011) 035
  [1109.6548].

\bibitem{sr}
  G.~Sigl and G.~Raffelt,
  {\em General kinetic description of relativistic mixed neutrinos,}
  Nucl.\ Phys.\ B {406} (1993) 423.

\bibitem{ars}
  E.K.~Akhmedov, V.A.~Rubakov and A.Y.~Smirnov,
  {\em Baryogenesis via neutrino oscillations,}
  Phys.\ Rev.\ Lett.\  {81} (1998) 1359
  [hep-ph/9803255].

\bibitem{as}
  T.~Asaka and M.~Shaposhnikov,
  {\em The $\nu$MSM, dark matter and baryon asymmetry of the universe,}
  Phys.\ Lett.\ B {620} (2005) 17
  [hep-ph/0505013].

\bibitem{Bulk_wdep}
  M.~Laine, A.~Vuorinen and Y.~Zhu,
  {\em Next-to-leading order thermal spectral functions
  in the perturbative domain,}
  JHEP {09} (2011)  084
  [1108.1259].
  
\end{thebibliography}
\end{document}